\def\lesssim{\mathrel{\hbox{\rlap{\hbox{\lower4pt\hbox{$\sim$}}}\hbox{$<$}}}}

\def\gtrsim{\mathrel{\hbox{\rlap{\hbox{\lower4pt\hbox{$\sim$}}}\hbox{$>$}}}}

\def\msun{M$_{\odot}$}

\def\teff{$T_{\rm eff}$}

\def\ll_lsun{log$({L/\rm L_{\odot}})$~}

\def\masa_msun{$M/ \rm M_{\odot}$~}

\def\m_mstar{$M/M_{*}$~}

\documentclass{aa}
\usepackage{graphicx}
        
\begin{document}

\title{Full evolutionary models for PG1159 stars. \\Implications for the
helium-rich O(He) stars.}

\author{ M. M. Miller Bertolami$^{1,2,3}$\thanks{Fellow of CONICET, Argentina.},L. G. Althaus$^{1,2}$\thanks{Member of the Carrera del Investigador
        Cient\'{\i}fico y Tecnol\'ogico and IALP, CONICET / FCAG-UNLP,
        Argentina.} }
\offprints{M. M. Miller Bertolami}

\institute{
$^1$   Facultad   de   Ciencias  Astron\'omicas   y   Geof\'{\i}sicas,
Universidad Nacional de La Plata,  Paseo del Bosque S/N, (B1900FWA) La
Plata, Argentina.\\
$^2$ Instituto  de Astrof\'{\i}sica La Plata, IALP, CONICET-UNLP\\
$^3$ Max-Planck-Institut f\"ur Astrophysik, Garching, Germany
\email{mmiller,althaus@fcaglp.unlp.edu.ar} }

\date{Received; accepted}

\abstract{}{We present full evolutionary calculations
appropriate to post-AGB PG1159 stars for a wide range of stellar
masses.} {We take into account the complete evolutionary stages of PG1159
progenitors starting from the Zero Age Main Sequence. We
consider the two kinds of Born Again Scenarios, the very late thermal
pulse (VLTP) and the late thermal pulse (LTP), that give rise to
hydrogen-deficient compositions. The location of our PG1159 tracks in
the effective temperature - gravity diagram and their comparison with
previous calculations as well as the resulting surface compositions are
discussed at some length.}{ Our results reinforce the idea that the
different abundances of $^{14}$N observed at the surface of those
PG1159 stars with undetected hydrogen is an indication that the
progenitors of these stars would have evolved through a VLTP episode, 
where most of the hydrogen content of
the remnant is burnt, or LTP, where hydrogen is
not burnt but instead diluted to very low surface abundances. We
derive new values for spectroscopical masses based on these new
models. We discuss the correlation between the presence of planetary
nebulae and the $^{14}$N abundance as another indicator that
$^{14}$N-rich objects should come from a VLTP episode while
$^{14}$N-deficient ones should be the result of a LTP. Finally, we
discuss an evolutionary scenario that could explain the existence of
PG1159 stars with unusually high helium abundances and a possible
evolutionary connection between these stars and the low mass O(He)
stars. }  {}


\authorrunning{Miller Bertolami \& Althaus}

\titlerunning{Full evolutionary models for PG1159 stars.}

\maketitle

%-------------------------------------------------------------------

\section{Introduction}

It is accepted that the hot, luminous PG1159 stars constitute the
immediate descendants of post-asymptotic giant branch (post-AGB)
evolution. They provide an evolutionary link between these stars and
most of the hydrogen (H)-deficient white dwarf (WD)
stars. Specifically, PG1159 are thought to be formed via a born again
scenario, that is, a very late thermal pulse (VLTP) experienced by a
hot WD during its early cooling phase (Fujimoto 1977, Sch\"onberner
1979 and Iben et al.  1983 for earlier references) or a late thermal
pulse (LTP) that occurs during the post-AGB evolution when H burning
is still active (see Bl{\" o}cker 2001 for references).  During the
VLTP, an outward-growing convection zone powered by helium burning
shell develops and reaches the H-rich envelope of the star, with the
consequence that most of the hydrogen content is burnt.  The star is
then forced to evolve rapidly back to the AGB and finally into the
central star of a planetary nebula at high effective temperatures
(\teff) but now as a H-deficient, quiescent helium-burning object. LTP
also leads to a hydrogen-deficient composition but as a result of a
dilution episode. Consequently, while post-VLTP models are
expected to become H-depleted (DB) white dwarfs (Althaus et
al. 2005a), it has been argued that post-LTP models may finally (after
the PG1159 stage) evolve into H-rich (DA) WDs with thin H
envelopes (Althaus et al. 2005b).

Currently, 37 stars are members of the PG1159 family, which span a
wide domain in the log g-\teff\ diagram: 5.5 $\lesssim$ log g
$\lesssim$ 8 and 75000 K $\lesssim$
\teff $\lesssim$ 200000 K (see Werner \& Herwig
2006 for a review). Roughly half of them are still embedded in a
planetary nebula.  A striking feature characterizing PG1159 stars is
their peculiar surface chemical composition. Indeed, spectroscopic
analyses have revealed that most of PG1159 stars exhibit H-deficient
and helium-, carbon- and oxygen-rich surface abundances. Surface mass
abundance of about 0.33 He, 0.5 C and 0.17 O are reported as typical,
though notable variations are found from star to star (Dreizler \&
Heber 1998; Werner 2001).  In fact, almost every star has its
individual He/C/O mixture.  According to Werner (2001) helium
abundances range between 33-85 \% and carbon and oxygen abundances
span the ranges 13-58 and 2-24 \%, respectively.  The rich variety of
surface patterns observed in PG1159 stars poses indeed a real
challenge to the theory of post-AGB evolution.  In particular, the
appreciable abundance of oxygen in the atmospheres of these stars has
been successfully explained by Herwig et al.  (1999) on the basis of
evolutionary calculations of the born-again scenario that incorporate
convective overshoot. In addition, hydrogen has been detected in some PG1159s
(called hybrid-PG1159 stars). These H-rich PG1159s are currently
interpreted in terms of a final helium shell flash experienced by the
progenitor stars immediately before their departure from the AGB (AGB
final thermal pulse or AFTP scenario, Herwig 2001).  Interest in
PG1159 stars is also motivated by the fact that about ten of them
exhibit multiperiodic luminosity variations caused by non-radial
$g$-mode pulsations, a fact that has attracted the attention of
researchers (see Quirion et al. 2004; Gautschy et al. 2005; C\'orsico
\& Althaus 2006 for recent works).

Needless to say, many relevant aspects about PG1159 stars, for
instance mass determinations, precise asteroseismological
fittings,  pulsation stability analysis and ages call for the need
of extensive full evolutionary models of PG1159 stars with different
stellar masses and realistic histories that led them through the
thermally pulsing AGB and born again phases (see Dreizler \& Heber
1998; Werner 2001). The lack of a grid of evolutionary sequences with
different masses in the literature has motivated us to undertake the
present work, which is aimed at providing full evolutionary tracks
appropriate for PG1159 stars based on a complete and selfconsistent
treatment of the progenitor stars, starting from the main sequence.
%We believe  that the evolutionary  models presented here  constitute a
%solid base to make sound predictions about for instance the dispersion
%in surface abundances expected in some PG1159 stars. 
Specifically, we follow the evolution of initially 1, 2.2, 2.5, 3.05,
3.5 and 5.5 $M_{\sun}$ stars from the zero-age main sequence (ZAMS)
through the thermally pulsing and mass-loss phases on the AGB to the
PG1159 regime. In particular, the evolutionary stages corresponding to
the almost complete burning of protons following the occurrence of the
VLTP and the ensuing born-again episode are carefully followed for
each sequence.  In addition we have followed the evolution of
initially 2.2 and 2.7 \msun\ model stars through the surface hydrogen
dilution phase that follows the LTP. Attention is paid to the
abundance changes during the entire evolution, which are described by
means of a time-dependent scheme for the simultaneous treatment of
nuclear evolution and various mixing processes. We also develop an
evolutionary scenario to address the existence of PG1159 stars with
unusually high helium abundances and nitrogen  deficiency, an
issue recently raised by Quirion et al. (2004) in the context of
pulsating PG1159s.  We also apply this evolutionary scenario
to address the existence of low-mass O(He) stars.

The paper  is organized  as
follows.  The  following section contains the main  physical inputs to
the  models and a description of the evolutionary sequences
considered. In  Sect.    3  we  present   the  evolutionary
results. There,  we elaborate on  the main aspects of  pre-PG1159
evolution, particularly during the  born-again phase and the attendant
chemical changes.   In this section we also compare our
PG1159 models with those existing in the literature.   In  Sect.  4,   
we  discuss  the
implications of our  results. We also discuss an evolutionary
scenario that could explain the existence of PG1159s with high
helium content. Finally, Sect. 5  is devoted to summarizing
our results.

%-------------------------------------------------------------------

\section{Input physics and evolutionary sequences}

The  calculations presented  in this  work  have been  done with  the
LPCODE evolutionary  code appropriate for  computing the formation and 
evolution  of WD
stars through late thermal pulses (Althaus et al. 2005a,b).
  LPCODE uses OPAL radiative opacities (including
carbon- and oxygen-rich compositions) from the compilation of Iglesias
\&  Rogers  (1996),  complemented,   at  low  temperatures,  with  the
molecular opacities from Alexander \& Ferguson (1994). In this work
we adopt solar metallicity. The abundance evolution of 16 chemical elements is
included via a time-dependent numerical scheme that simultaneously
treats nuclear evolution and mixing processes due to convection, salt
fingers and overshooting (OV). Such a treatment is particularly important
during the short-lived phase of the VLTP and the ensuing born-again
episode.  Computationally, convective OV was treated as a
exponentially decaying  diffusive process above and below any formally
convective region, including the convective core (main sequence and
central helium burning phase), the helium flash convection zone and the 
external convective envelope (see
Herwig et al. 1999 for details). 
\begin{table*}[ht!] 
\centering
\begin{tabular}{cccccc}\hline
Initial Mass & $\Delta t_{\rm MS}$ & $t_{\rm CHeB}$ & $t_{\rm TP-AGB}$
& \#TP & final mass [in \msun]\\ \hline 
1 \msun\ & 9790 & 12666-12790 & 12804 & 5/7 & 0.530/0.542 \\ 
2.2 \msun\ & 879 & 904-1158 & 1183 & 16 & 0.565 \\ 
2.5 \msun\ & 595 & 609-818 & 832 & 13 & 0.584 \\ 
3.05 \msun\ & 374 & 380-473 & 480 & 11 & 0.609 \\ 
3.5 \msun\ & 254.5 & 257.5-314 & 319 & 8 & 0.664 \\
 5.5 \msun\ & 81 & 81.6-92.5 & 93.3 & 8 & 0.870 \\
\end{tabular} 
\caption{Times for different moments of the evolution of the PG1159 
progenitors (in $10^6$ yr since the ZAMS). The column below $t_{\rm CHeB}$ 
marks the times at the start and the end of core helium burning (CHeB). 
Also the number of thermal pulses is given (column 5).}
\label{tab:previa} 
\end{table*}

We present six model star sequences, with initially 1, 2.2, 2.5,
3.05, 3.5, 5.5 \msun. All of the sequences are followed from the
ZAMS through the thermally pulsing and mass
loss phases on the AGB (TP-AGB). After experiencing several thermal
pulses (see Table \ref{tab:previa} for a description of the evolution
of the sequences up to the TP-AGB), the progenitors depart from the
AGB and evolve towards high \teff. Mass loss during
the departure from the AGB has been arbitrarily set, until log($T_{\rm
eff})\sim4$ when it was turned off, as to obtain a final helium shell
 flash during the early WD cooling phase.
 
We mention that the 1 \msun\  model experienced a core helium flash, which
led it to the horizontal branch to burn helium in a stable
fashion. For this sequence two different AGB evolutions have been 
considered, with different mass loss rates as to obtain different
number of thermal pulses and, eventually, two different remnant masses
of 0.53 and 0.542 \msun. After mass loss episodes, the stellar
masses of the post-AGB models is reduced to 0.53, 0.542, 0.565, 0.584,
0.609, 0.664 and 0.87.
\begin{figure*}
\centering
\includegraphics[clip,width=17 cm]{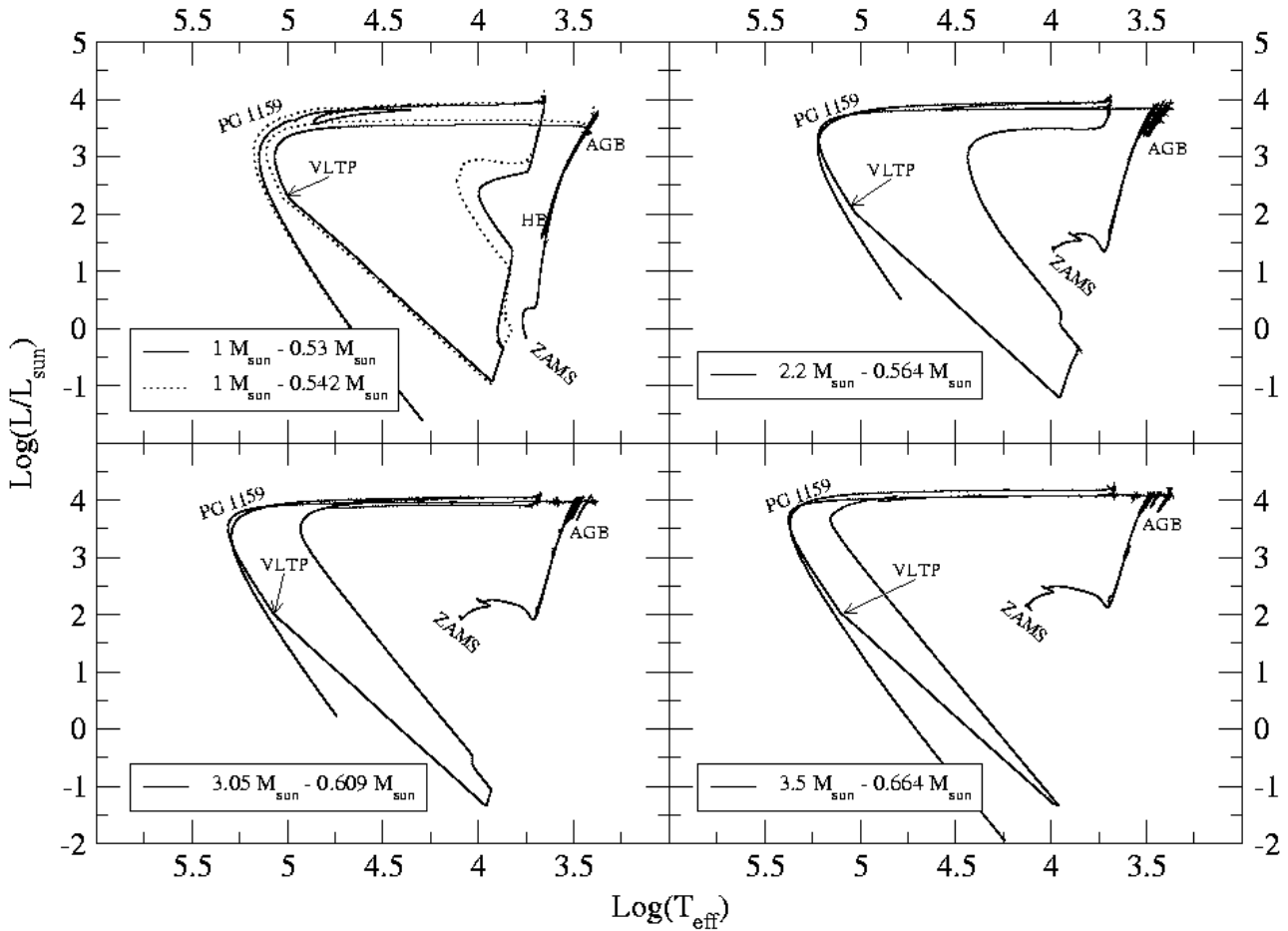}
\caption{HRDs for selected VLTP sequences. Legends indicate initial and final 
masses of the models.  In the upper left panel, the dotted line corresponds
to the case in which a larger number of thermal pulses was allowed
before the progenitor departs from the AGB. }
\label{fig:HRS}
\end{figure*}
Except for the 1 \msun\ sequences we find convergence problems shortly
after the departure from the AGB, probably due to the lack of a
hydrostatic solution in our models (see Wood \& Faulkner 1986);
problems that are more noticeable when OPAL opacities are used.  To
overcome this difficulty we have temporarily increased the mixing
length theory parameter $\alpha$ by about a factor of two during the
early departure from the AGB (log $T_{\rm eff}\lesssim 3.6$).  For the
purpose of the present work, this artificial procedure does not bear
any relevance for the further stages of evolution. We also want to
mention that our  most massive model (0.870
\msun) experienced several convergence problems, in the envelope and
atmosphere integration, due to the very high radiation pressure 
 that characterizes the outer layers of this model. This causes the
density of the hydrostatic model to drop to almost zero and thus our
sequence failed to converge. To continue with evolution,  we
imposed artificial boundary conditions during its redward excursion
after the VLTP. We particularly checked that this modification did not alter 
the location of the model in the  log g-\teff\ diagram  by doing the same 
modification to
the 3.5 \msun\ model, for which the evolution after the VLTP was
consistently calculated, and we find that no effect of this artificial
evolution was present when the model reached the PG1159 phase.

\section{Evolutionary results}

\subsection{Evolution during the born-again phase}
\begin{figure}
\centering
\includegraphics[clip,width=250pt]{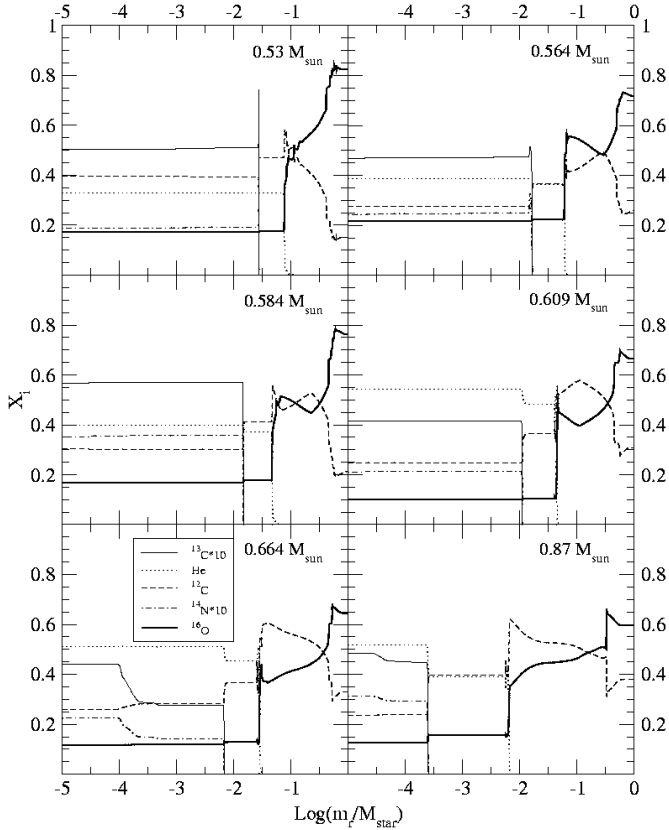}
\caption{Internal chemical profiles for VLTP models having 
different stellar masses shortly after the star has attained high luminosities 
after the VLTP. These chemical profiles will be affected by the  
envelope mixing and mass loss episodes when models return to the 
domain of the giant region.  Abundances are given in mass fractions.}
\label{fig:perfiles}
\end{figure}
In this work we do not intend to present a detailed study of the born
again phase leading to PG1159 stars, an aspect which will be deferred to
a forthcoming publication. Instead we limit our discussion to
presenting the main characteristics of the evolution of the progenitor
stars relevant for PG1159s. In Fig. \ref{fig:HRS} we show the
complete Hertzsprung-Russell diagram (HRD) for four of our
calculated sequences (for a complete discussion of the evolution
of an initially  2.5 \msun\ sequence see
Miller Bertolami et al. 2006). As can be noted, our numerical simulation
covers all the evolutionary phases from the ZAMS to the domain of the
early-WD including the stages corresponding to the helium thermal
pulses at the tip of the AGB and the born again episode. This last stage is particularly
fast and evolution proceeds on a scale of years (Herwig 2001,
Miller Bertolami et al. 2006). During this stage the star is ballooned
back to the giant region both by the helium flash and
by the violent H-burning that occurs at the VLTP. Note that the detailed
evolution of the sequences during this stage depends on the mass of
the remnant. In particular the expansion caused by H-burning is very
dependent on the mass of the model. We find that the double loop path
in the HR, characteristic of successive H- and He-driven expansions
(Lawlor \& MacDonald 2003), is only present for intermediate mass
models ($0.55$\msun$\lesssim M \lesssim 0.61$\msun). This is because for model
with lower masses the H-driven expansion is strong enough to keep the
model at low  effective temperatures until the effect of the He-driven
expansion takes over, so that no return to the blue after the
H-driven expansion happens. On the other hand, for massive models we
find that H-driven expansion is extremely weak and it is almost
suppressed (it stops at log($T_{\rm eff})\lesssim 4.75$), see
Fig. \ref{fig:HRS}.

Another point worth of comment, without entering  a detailed
description of the born again scenario, is that the effective
temperature at which the model emerges at high luminosities is a
function of the stellar mass. As can be seen in  Fig.  \ref{fig:HRS} we find 
that the lower the
mass of the model, the lower the effective temperature at which the
model becomes bright (from log($T_{\rm eff})\sim 4$ for the 0.53
\msun\ to 5.1 for the 0.664 \msun).
\begin{figure*}
\centering
\includegraphics[clip,width=14 cm]{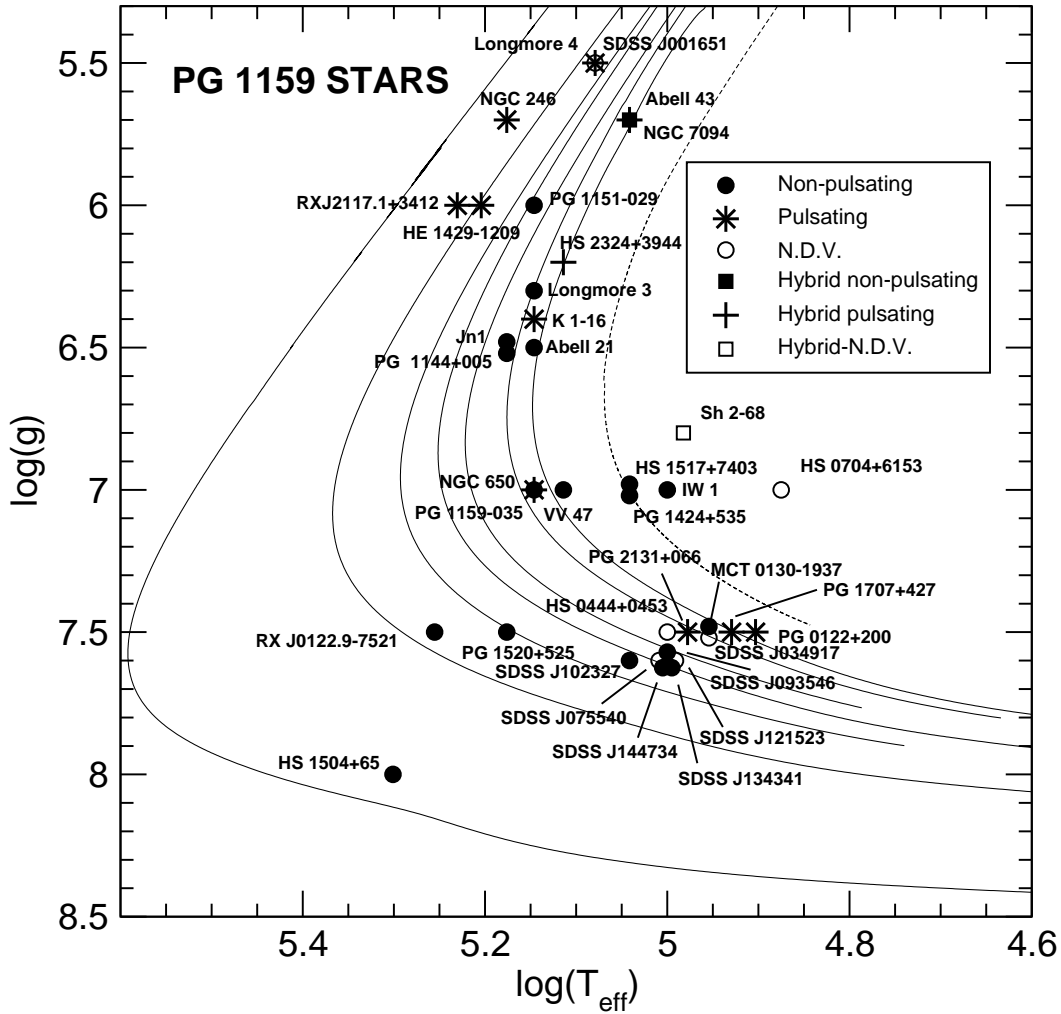}
\caption{ Location of  analyzed PG1159 stars ( values taken from Werner \& Herwig 2006). Empty symbols are used 
for the objects for which no information about their variability is
available (N.D.V.).  Solid lines correspond to our evolutionary
sequences 0.87, 0.664, 0.609, 0.584, 0.564, 0.542 and 0.53
\msun. The broken line corresponds to the (helium enriched) 0.512\msun\ post
early AGB model discussed in Section 4.2.}
\label{fig:PG1159s}
\end{figure*}

In Fig.\ref{fig:perfiles} we show chemical abundances (abundances
throughout this article are  given in mass fractions) 
throughout the interior of most of the models presented in this work
(including the previously calculated 2.5 \msun) at the moment 
when the models have attained high luminosities after the VLTP. Note
the dependance of the composition profile on the stellar mass. The
different envelope compositions are due partly to the fact that each
sequence underwent a different number of thermal pulses (see Table 1)
during the AGB phase\footnote{ We find that the intershell helium abundance reaches a
local minimum value after the third-seventh (depending on stellar
mass) thermal pulse (see also Fig. 11 in Herwig 2000).}. The
abundance profiles shown in Fig.\ref{fig:perfiles} correspond to
evolutionary stages well before the occurrence of the envelope mixing
by convection and mass loss episodes that take place at very low
temperatures after the VLTP. Consequently the depicted surface
abundances of these models $are$ $not$ strictly the ones expected in
PG1159s \footnote{Because these abundances should be the consequence
of both the envelope dilution and mass loss. We note that dilution of
the envelope material with the intershell layers leads to lower
$^{14}$N surface abundances, between 0.005 and 0.025 by mass which is
similar to the observed ones (Miksa et al. 2002).  }. As can be seen
in Fig.\ref{fig:perfiles} after the VLTP models show a chemical
structure that can be summarized as follows: a C-O core, an intershell
region rich in He, C, O and a surrounding envelope where also high
amounts of $^{13}$C and $^{14}$N (created at the VLTP) are found. We
elect to show the chemical stratification prior to the envelope mixing
and mass loss episodes because both the deepness of the envelope
convection after the VLTP and the mass loss rate are highly uncertain
and will affect the final surface composition. For example all of our
models remain at log$(T_{\rm eff})\lesssim 3.8$
\footnote{Except for the 0.870 \msun\ for which we cannot
obtain reliable ages.} for about 700 - 800 yr. Mass loss rates similar
 to those observed in the Sakurai's object
 ($1\times10^{-5}$\msun/yr, and even higher, see Hajduk et al.
 2005) would cause $7\times10^{-3}$
\msun\ to be stripped off from the star
\footnote{This  strong mass loss at low effective temperatures may help to explain the bizarre surface abundance of H 1504+65, which is difficult to explain  by only considering mass loss during the standard post-AGB evolution (Sch\"onberner \& Bl\"ocker 1992).}. In all the cases the expected surface abundances should be
somewhat between those displayed by models in the envelope and at the
intershell region (Fig. \ref{fig:perfiles}). The helium abundance is
particularly noteworthy. Indeed note that our post-VLTP models predict
that the final surface abundance of helium will be in the range
0.3-0.55 by mass, which is in agreement with the range of observed
helium abundance in most PG1159s (see Werner \& Herwig 2006).  However
these abundances should be taken with caution as they are not only
determined by the stellar mass but also by the number of pulses
considered in the AGB stage and the exact value of the adopted
OV parameter ($f$) (Herwig 2000). 

% The high helium abundance
%we found in some cases bears some relevance in the light of what have
%recently been proposed by Quirion et al.(2004) that the helium
%enrichment in the envelope plays a key role in explaining the
%existence of non-pulsating PG1159s in the instability strip.
%

%
\subsection{PG1159 stage: Overall location of our models in the 
log g-\teff\ diagram}

\begin{figure}
\centering
\includegraphics[clip,width=250pt]{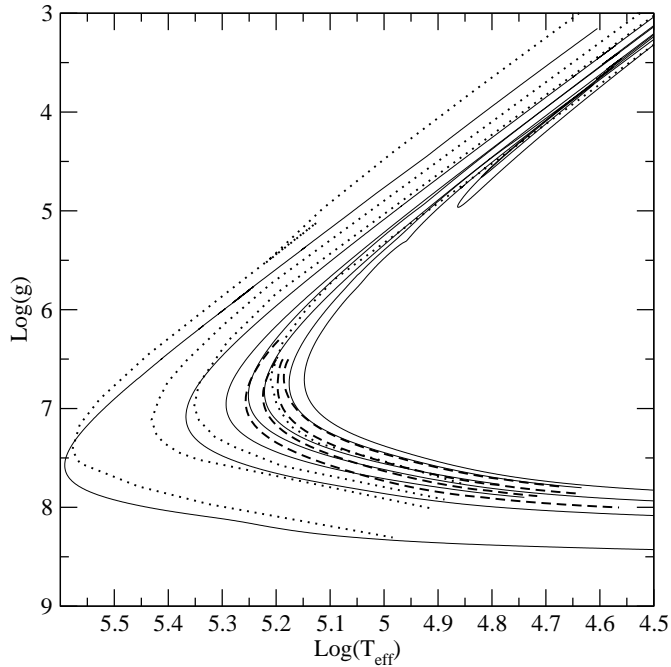}
\caption{Comparison of the models presented in this work (thin lines; 0.87, 0.664, 0.609, 0.584, 0.565, 0.542, 0.53 \msun) with previous models of He-burning objects. Dotted lines correspond to Wood \& Faulkner 1986 type B mass loss models (0.87, 0.76, 0.7, 0.6 \msun) and broken lines indicate O'Brien \& Kawaler tracks for models with PG1159 like compositions (0.63, 0.6, 0.573, 0.55 \msun). As can be seen our models are generally hotter than previous models of similar mass.  }
\label{fig:wood-obrien}
\end{figure}

In this section we focus on the location of the models in the
log g-\teff\ diagram. We
compare them with the  available data from observations and from
previous stellar models for H-deficient post-AGB stars (Figs. \ref{fig:PG1159s}
\& \ref{fig:wood-obrien} respectively).  As can be seen in Fig. 
\ref{fig:wood-obrien} our models are somewhat hotter than previous
models of He-burning objects with H-deficient surface abundances.
From visual inspection one can conclude that the mean shift in
spectroscopical masses due to this difference in the surface
temperatures would be 0.025 \msun\ for O'Brien \& Kawaler (as
presented in Dreizler \& Heber 1998) and 0.04 \msun\ for Wood \&
Faulkner models (Wood \& Faulkner 1986).  On the other hand, we find
our models to be in good agreement with the 0.604 track presented by
Werner \& Herwig (2006), which is based on similar physical
assumptions. This  seems to confirm the finding by these authors
that the location of the models in the log g-\teff\ diagram is very
sensitive to the modeling of the previous AGB evolution. In
particular Werner \& Herwig (2006) point out that this difference may
be related with the effect of a smaller effective core growth caused
by the more efficient third dredge-up events of the new models (mainly
as consequence of the inclusion of overshooting mixing at the He-flash
driven convection zone during the TP-AGB). If this is true, a lower
$f$-value than adopted in the present article ($f=0.016$) at the
He-flash driven convection zone would lead to a less efficient third
dredge-up and, thus, would shift post-AGB tracks to lower effective
temperatures\footnote{After the acceptance of this paper we have not
found (in preliminary calculations) any important shift for remnants
of similar mass but different effective core growths during the
TP-AGB. We intend to explore this effect in a future work}.  This
speculation would be in line with hydrodynamical simulations of the
He-shell flash (Herwig et al. 2006) that suggest that OV may be
reduced to a minimum. However, it should be kept in mind that a quite
efficient overshooting at the base of the shell flash driven
convective zone (during the TP-AGB) is needed in order to obtain
oxygen surface abundances like those displayed by PG1159 stars (Herwig
et al. 1999).

\begin{figure}
\centering
\includegraphics[clip,width=250pt]{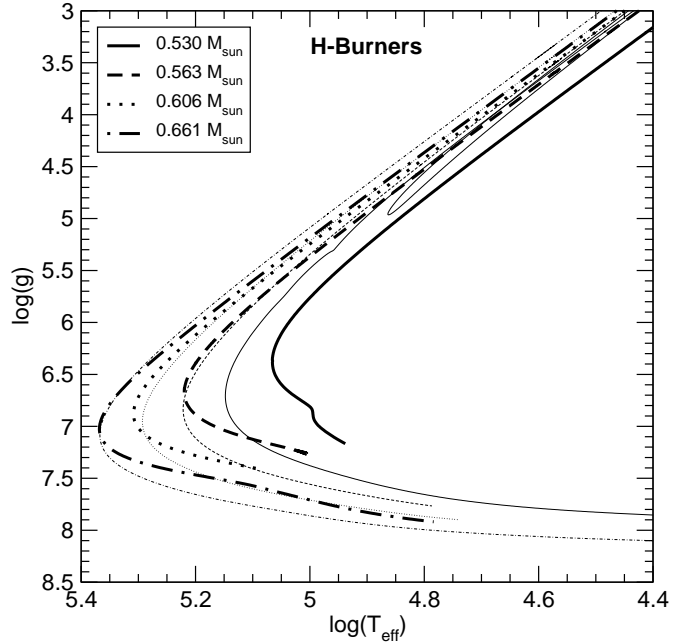}
\caption{Comparison of the location of H and He burning post AGB models. 
Note that at  lower g-values (during the WC stage) H-burner are located 
at slightly larger gravities, while at higher gravities (beyond 
log $g\sim6.5$) H-burners have lower gravities than He burning objects. 
Also note that the low mass H-burning object lies at lower temperatures 
than the He-burning object of similar mass (0.53 \msun).}
\label{fig:Hburner}
\end{figure}

%\begin{figure}
%\centering
%\includegraphics[clip,width=250pt]{T-g-contutti.eps}
%\caption{Comparison with observationally derived parameters for  H-deficient stars. We show data for PG1159 stars and their probable progenitors WC stars (see Werner 2001 for references).  }
%\label{fig:t-g contodo}
%\end{figure}
%
 Also regarding the location of post-AGB tracks in the log g-\teff\
diagram, previous works (Werner 2001) have noted that it was possible
that spectroscopical masses which are based on models with normal
surface abundances may suffer from systematic errors. As it is shown
in Fig.
\ref{fig:Hburner} H-burning models are indeed different from those
which are He-burning of similar mass. However the effect is not the
same for all masses. We find that differences are larger for models of
lower mass. In fact our H-burning model of 0.53 \msun\ is located at
much lower temperatures than the He-burning model of similar mass (and
thus closer to the location of the low mass models of Bl{\" o}cker
1995).  Another difference is that in the high-g side of
the knee, in the log g-\teff\ diagram, H-burning models are
systematically less compact.

\section{Discussion}
\subsection{Are PG1159  stars post VLTP or LTP stars?}
\begin{figure}
\centering
\includegraphics[clip,width=250pt]{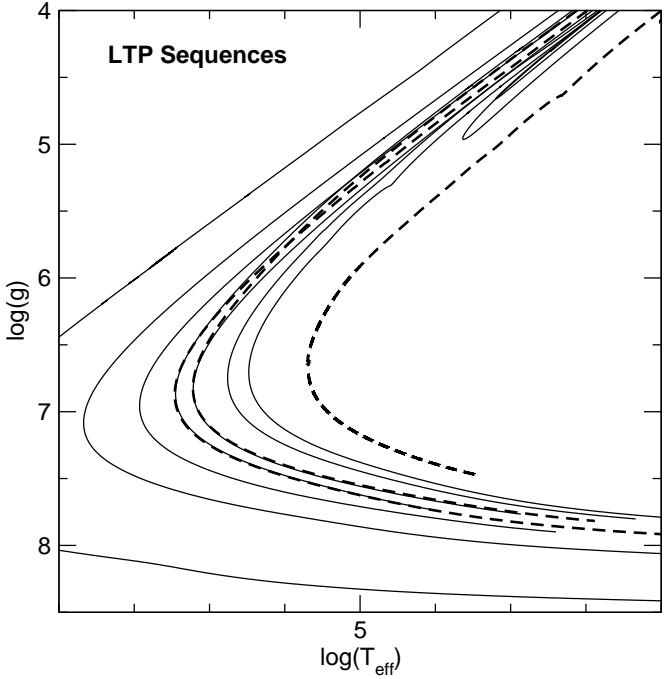}
\caption{LTP sequences calculated for this work (0.589 \msun, 0.564 \msun, broken lines) are shown agains VLTP sequences (thin lines). Also, the helium
enhanced post early AGB sequence with 0.512 \msun\ (which is also a
post-LTP model) is shown (coolest sequence, also in broken
lines). Note that LTP and VLTP models of similar mass follow almost
the same track.}
\label{fig:LTP-VLTP}
\end{figure}

It is worth mentioning that all of our VLTP-models display high
amounts of $^{14}$N at their surface (and also a high
$^{13}$C/$^{12}$C ratio see Fig. \ref{fig:perfiles}).  This is
similar to the result of Iben \& MacDonald (1995), but at variance
with later results for the VLTP by Herwig (2001) . Since many PG1159s
do not exhibit $^{14}$N (Dreizler \& Heber 1998), its was interesting
to analyze other possible channels for the formation of PG 1159 stars
which do not yield high $^{14}$N surface abundances. In particular it
was interesting to look for changes in the location of the models in
the log g-\teff\ diagram if PG1159 models were the offspring of a LTP,
in which H is not burnt but diluted after the helium flash and thus no
enhanced $^{14}$N abundance is expected. Consequently we carried out
two additional calculations of sequences that went through a LTP (of
0.589 and 0.562
\msun). These are shown in Fig. \ref{fig:LTP-VLTP} where it can be seen
their location in the log g-\teff\ diagram compared with VLTP
sequences of similar mass.  As can be noted there is only a slight
unimportant difference between VLTP and LTP tracks that occurs in the
low-gravity region before reaching the knee\footnote{ However
while our models are calculated adopting solar values for the heavy
metal abundances, it is known that post-VLTP stars display a non-solar
heavy metal distribution (Asplund et al. 1999 \& Miksa et al. 2002).
For example, we note that tracks move 0.03 dex to the right (red) if
opacities are increased by $\sim40$\%.}.   This may be connected
to the fact that, there, the H-burning shell is still active in LTP
sequences.  
%In what follows we will discuss the implications of our
%evolutionary sequences in the light of the correlation between the
%presence of $^{14}$N and pulsations found by Dreizler \& Heber
%(1998). In fact, we will assume that this correlation is valid for all
%PG1159 non-pulsators within the instability strip (however see Werner
%\& Herwig 2006 for a redetermination of nitrogen, 0.001 by mass, in
%the pulsating PG1159-035), even though this correlation is based on a
%very small sample (9 stars).

\begin{figure}
\centering
\includegraphics[clip,width=250pt]{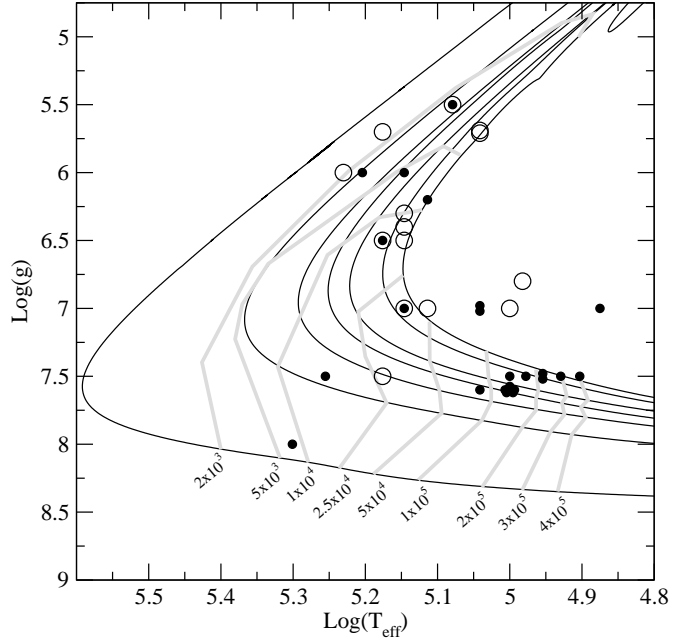}
\caption{Isochrones for post-VLTP sequences. Times are measured since the last helium flash. Empty circles (black dots) indicate PG1159 stars with (without) planetary nebulae. Note the existence of some objects beyond the $2.5\times10^4$ 
yr isochrone that show planetary nebulae. These objects could not be
the result of a VLTP episode as their surrounding nebulae should have
faded. Indeed, in none of these objects significant amounts of
$^{14}$N (a signature of previous proton burning during the VLTP) have
been found while many of them are known to be $^{14}$N-deficient
(Dreizler \& Heber 1998). Also note the fast evolution of the  most
massive model before it reaches log g$\sim 8$, which easily explains
the absence of  detected PG1159 stars with log T$_{\rm eff}>$5.4}
\label{fig:iso}
\end{figure}

In both LTP-sequences, H was diluted to about a 4\% in
mass\footnote{He/C/O abundances are 31/40/20 and 33/36/21 for
the 0.589 and
0.562 \msun models, respectively.}, which is close to the detection limit
(Werner 2001). Our  LTP-models do not display enhanced $^{14}$N
abundance at their surface. 
%This means that $^{14}$N deficiency in 
%non-pulsators
%might be taken as a signal of the presence of hidden H in the envelope
%of these objects (however see Section 4.2).  
Our results reinforce the idea of Werner et al. (1998), that the
different abundances of $^{14}$N in the PG1159 sample seems to be an
indication that PG1159 progenitors are a mixture of VLTP and LTP-AFTP
cases.  As was mentioned by Werner (2001) the existence of LTP
progenitors in the PG1159 sample may help to solve the inconsistency
presented by the expansion ages of planetary nebulae. In this connection we find it interesting to discuss the
presence of planetary nebulae regarding the location of their central
stars in the log g-\teff\ diagram and their corresponding evolutionary
ages. In Fig. \ref{fig:iso} we draw isochrones based on the
evolutionary ages of our models. We note that ages are measured since
the moment of the VLTP and not since the first departure from the
AGB\footnote{Also note that, as no mass loss has been considered in
these sequences after the VLTP, these values should be taken as upper
limits.}.

  In order to compare theory with observations, and to distinguish
between stars coming from VLTP, LTP and AFTP scenarios, it seems
necessary to see how observable parameters like, for example, the presence of
planetary nebulae, surface abundances and pulsational instability
correlate with each other. In this context it is interesting to see
how the presence of high $^{14}$N surface abundances correlates with
the existence of planetary nebulae around PG1159 stars. In what
follows we will discuss the implications of our evolutionary sequences
in the light of the correlation between the presence of $^{14}$N and
pulsations found by Dreizler \& Heber (1998). In fact, we will assume
that this correlation is valid for all PG1159 non-pulsators within the
instability strip (however see Werner
\& Herwig 2006 for a redetermination of nitrogen, 0.001 by mass, in
the pulsating PG1159-035), even though this correlation is based on a
very small sample (9 stars). Although this is very speculative, we
think that inferences may be interesting enough. Also note that,
although $^{14}$N has been demonstrated by Quirion et al. (2004) to be
unimportant in driving pulsations,$^{14}$N is a witness of previous
evolution. Stars with and without $^{14}$N are expected to have
different H and heavy elements abundances and consequently different
opacities (see Iglesias et al. 1995).  As the first travel to the
high temperature region of the HR (after departing from the AGB) takes
about 10000-20000 yr (VLTP, Bl{\" o}cker 1995) for low mass stars, we
find it interesting to note that the presence of planetary nebulae 
around some PG1159s close to the isochrones of 25000 and 50000 yr
clearly implies that these objects (particularly NGC 650, VV47 and
PG1520+525) cannot be the offspring of a VLTP (because planetary
nebulae should have faded during this time interval, 30000-75000 yr).
This speculation seems to be corroborated by the fact that one of
these objects (PG1520+525) is known to be $^{14}$N-deficient, and also
by the fact that the rest of them are non-pulsating objects which are
close to, or inside, the instability strip (and thus according to
Dreizler \& Heber 1998 must be $^{14}$N-deficient).  It is also worth
mentioning that some stars (between the 25000 and 100000 yr isochrone)
with detected nitrogen (the ones that should be coming from a VLTP
episode, for instance PG1144+005) do not show an associated planetary
nebulae, whilst NGC 650 and VV45 (both non-pulsating objects) do show
planetary nebulae. This correlation between the presence of $^{14}$N
and the absence of the planetary nebulae, seems to be another
indicator that $^{14}$N-rich objects should come from a VLTP while
$^{14}$N-deficient should come from a LTP.

This argument  is not affected by the possibility that the
pulsating PG1159 at relatively low g ($\sim6$)  (which do
display planetary nebulae) may be $^{14}$N rich according to the
correlation found by Dreizler \& Heber (1998). In fact, because all
these objects are more massive than $\sim$ 0.6 \msun\ according to our
models, then these objects quickly reach the WD cooling track
where they experience the VLTP ($\sim 5000$ yr, see Bl{\" o}cker 1995). Then,
as these objects are still close to the 2000 yr isochrone, no more than
$\sim7000$ yr have past since their first departure from the AGB and
thus they can still retain their planetary nebulae, even if they have
suffered from a VLTP.

 Even beyond the speculation about the correlation between $^{14}$N
and pulsations, it is interesting to note that PG1144+005 and
PG1159-035, which do not show associated planetary nebulae 
(while  in the log g-\teff\ diagram they are surrounded 
by stars that do show
associated planetary nebulae) are known to display $^{14}$N at their
surfaces. On the other hand, PG1520+525 which is located closer to the
same isochrone than PG1159-035 and that is known to be $^{14}$N
deficient (N$ \lesssim 1\times10^{-4}$, Miksa et al. 2002),
does show a planetary nebulae.

In this section we mentioned that the existence of two groups of
$^{14}$N-abundance in PG1159 stars could be explained by the existence
of two different evolutionary channels leading to the PG1159 stage
(namely VLTP and LTP episodes).  If so, our models predict
$^{14}$N-deficiency to be directly related with the presence of H
below the detection limit. We also mentioned that this claim might be
supported by the correlation between the presence of $^{14}$N and the
absence of planetary nebulae. However, it is clearly at variance with
the reported correlation between helium and $^{14}$N mentioned by
Quirion et al. (2004), as there is no simple reason why LTP models
only should produce helium-enriched PG1159 stars. In the following
section we will discuss an evolutionary scenario for these
helium-enriched stars (namely HS 1517+7403, HS 0740+6153 and MCT
0130-1937), an explanation which is related to the low mass of these
stars.

\subsection{Low T$_{\rm eff}$-Low $g$ PG1159 stars. The existence
of helium-enriched PG1159 stars  and their possible connection with 
low mass O(He) stars} 
\begin{figure}
\centering
\includegraphics[clip,width=250pt]{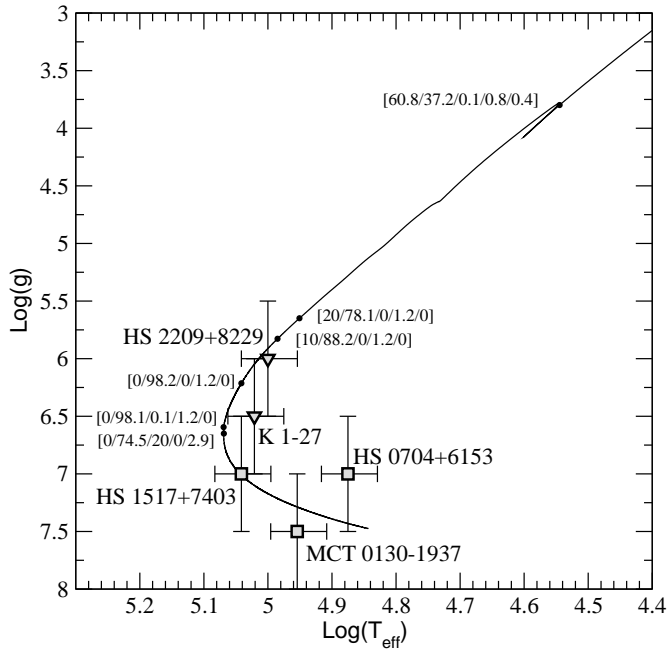}
\caption{Evolution in the log g-\teff\ diagram for the 0.512 \msun\ remnant. Surface abundances [H/He/$^{12}$C/$^{14}$N/$^{12}$O] diplayed by the model are given at certain points (black dots) for comparison with the abundances displayed by the low mass O(He) stars and the He-enriched PG1159 stars (triangles
and square symbols, respectively).}
\label{fig:ELTP}
\end{figure}
In the previous section we assume the fact that pulsating PG1159s are
rich in $^{14}$N; however as was noted by Dreizler \& Heber (1998) and
also by Quirion et al. (2004) many of the non-pulsator stars in the
sample have an unusually high helium abundance. This is important in
view of the results found by Quirion et al. (2004) that this helium
enrichment would be the cause for the existence of non-pulsators
within the instability strip\footnote{Note, however, that many
non-pulsators in the observational instability strip are found to have
standard helium abundances (Longmore 3, Abell 21, VV47, PG1424+535 and
PG1151-029).  On the other hand, NGC 246 which displays a high
helium abundance ($\sim60\%$, Werner \& Herwig 2006) is a pulsating
PG1159.}. If non-pulsators are indeed characterized by
helium-enhanced, N$^{14}$-deficient surface abundances they cannot be
the result of the standard LTP scenario experienced by post TP-AGB
stars.  Also, it is clear from visual inspection of
Fig. \ref{fig:PG1159s} that many of the H-deficient stars lie too much
to the red of the post-VLTP models presented in this work. Although
comparison with observations has to be taken with care due to high
uncertainties in the determination of $g$ (typical errors in log $g$
are about $\sim 0.5$ dex), we feel that this discrepancy between
models and real stars deserves some discussion.  Firstly, it is
interesting to note that these objects are not probably VLTP
descendants, as three of them (PG 1424+535, HS0704+6153 and HS
1517+7403) are known to be $^{14}$N-deficient while no
$^{14}$N-enhanced PG1159 is known in that region of the log g-\teff\
diagram.  Also as it is shown in Fig. \ref{fig:iso} some of these
objects still retain a planetary nebula, a fact that is difficult to
understand within the VLTP scenario (Werner 2001). TP-AGB progenitor
stars that went through a LTP is a possibility to explain the
existence of these PG1159s. However, as it is shown in
Fig. \ref{fig:LTP-VLTP}, LTP and VLTP models of similar masses seem to
be located at similar places in the log g-\teff\ diagram.  Thus if
these stars are descendents of LTP episodes they should be objects
with very low stellar masses\footnote{However it should be noted that
stars with very low mass  may need an unacceptable long time to
reach the PG1159 stage ($< 12.8$ Gyr). }.

We have explored an evolutionary scenario that could explain the
existence of the above mentioned PG1159s.  In addition we have found
that this scenario may offer an alternative explanation for the
formation of low mass O(He) stars. This subclass of H-deficient stars
is characterized by the fact that their members display almost pure
He-envelopes with only small traces of additional elements. They have
been proposed to be the offspring of a merging event between two white
dwarfs (Rauch et al. 2004, see also Saio \& Jeffery 2002 for the
simulation of the merging event) and the direct descendants of RCrB
stars. However at first glance it would seem that the final masses of
the products of a merging event (Saio \& Jeffery 2002) should be
higher than the masses derived for the two low mass O(He) stars
(roughly $\sim 0.55$\msun). In what follows we will show that the two
O(He) stars with low mass (namely \mbox{K\,1-27} and HS2209+8229) may be the
direct progenitors of the He-enriched PG1159  stars. In
particular it is worth noting that K1-27, for which detailed
abundances have been derived, displays a surface composition typical
of the He buffer of AGB stars, this is almost pure He ($\sim 98\%$)
with traces of $^{14}$N ($\sim 1.7\%$) (Werner \& Herwig 2006). 
Also both of these stars are supposed to have still an appreciable mass
loss rate of about $\sim 10^{-9}$\msun/yr, as derived from radiation
driven wind theory (see for example Table 4 of Rauch et al. 1998).

 In order to analyze an evolutionary channel that could give rise
to $^{14}$N-deficient, He-enhanced PG1159 stars, we have calculated
one additional sequence for an initially 1 \msun\ star, but changing
mass loss rates for it to suffer its first thermal pulse while it was
departing from the AGB (so its first thermal pulse is a LTP). This
situation is more likely to occur in low-mass star progenitors (see
Bl\"ocker 1995).  The resulting 0.512 \msun\ remnant is shown in
Fig. \ref{fig:PG1159s} and \ref{fig:LTP-VLTP} (post early
AGB sequence).  As can be seen this model can account for the location
of many of the ``redder'' PG1159 stars. However due to the low
intensity of the first thermal pulses in low-mass stars no strong
dredge-up is found and thus the sequence displays a normal surface
abundance composition after the LTP . We analyzed the possibility that
the model can turn into a PG1159 due to mass loss. This is an
important point as the model remains in the red giant region (where
strong mass loss episodes are suppossed to occur) for an unusually
long period of time after the LTP. To this end we follow the
evolution using a mass loss rate of $10^{-6}$
\msun /yr if log(T$_{\rm eff})<3.8$ and $10^{-8}$ \msun /yr if
log(T$_{\rm eff})>3.8$. We find that such mass loss rates after the
LTP can indeed turn the star into a PG1159. The fact that the model
remains in the red giant region for a long period of time is due to
the several He-subflashes it experiences after the first thermal pulse
(this sub-flashes are typical of the first thermal pulses of low-mass
stars). It should be noted that, as only one thermal pulse has
happened, the  final He/C/O abundances of this model (73/21/3) are much
different from typical PG1159 models.  Also as no vigorous H burning
happened, the star is also $^{14}$N-deficient. In this connection it
is interesting to note that the three He-enriched (and
$^{14}$N-deficient) objects in the non-pulsating sample of Dreizler \&
Heber (1998)\footnote{These are HS 0704+6153, HS 1517+7403 and MCT
0130-1937 (He/C/O= 69/21/9, 84/13/2, 74/22/3 respectively, Miksa et
al. 2002).} are located close to this track in the log g-\teff\
diagram and show a very similar composition. Indeed we think that the
mentioned correlation between He and $^{14}$N may only reflect that
the three non-pulsators with higher He-enriched surface abundances are
the offspring of low mass stars that never reached the TP-AGB and
suffered their first thermal pulse as a LTP.  

 In Fig. \ref{fig:ELTP} we show the location of the 0.512 \msun\
sequence in the log g-\teff\ diagram. Additionally the surface
abundances [H/He/$^{12}$C/$^{14}$N/$^{12}$O] displayed by the model
are given for certain points during its evolution. The first point (at
log \teff $\approx$ 4.6 denotes the moment in which surface abundances start
to change as consequence of mass loss. At this point layers in which
partial H-burning has happened are shown by the model. Abundances
continue changing until the small He-buffer (similar to that of case 3
in Iben 1995) appears at the surface. At this moment, and until the He
buffer is eroded by mass loss, the surface abundance of the models is
of 98 \% of He and 1.2 \% of $^{14}$N. These abundances are
(surprisingly) similar to those exhibited by K1-27, which is located at the
same region (see Fig.\ref{fig:ELTP}) like the model in the log
g-\teff\ diagram. Finally, when the He-buffer is removed by additional 
mass loss
the stars diplays surface abundances typical of the He-enriched
PG1159. In fact this happens while the model is passing through the
region (in the log g-\teff\ diagram) in which He-enriched PG1159 are
located. Although the exact locus in the log g-\teff\ diagram in which
the star will show O(He)- or He-enriched- surface abundances depends on
the exact value of the mass loss rate (which has been set 
somewhat artificially
in this work), this numerical experiment shows that it is possible
(with typical mass loss rates) that the low mass O(He) stars and the
He-enriched-PG1159 form an evolutionary sequence.

\section{Derivation of new spectroscopical masses}
\begin{figure}
\centering
%\raggedleft
\includegraphics[clip ,width=190pt,angle=-90]{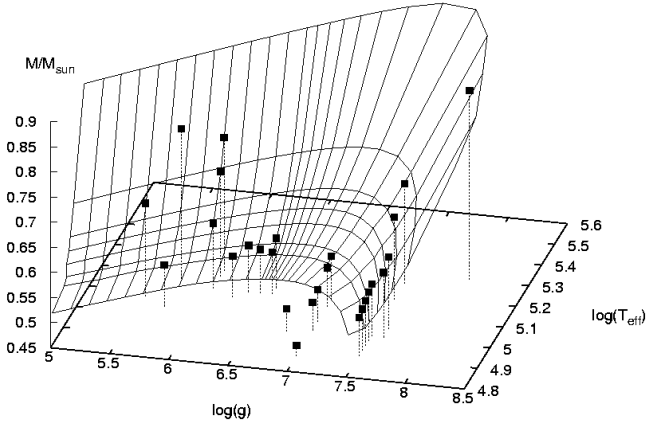}
\caption{Location of the sequences in the g-\teff-M space. Black squares show the derived location for real PG1159 stars by adopting the \teff-g values presented by Werner \& Herwig (2006).}
\label{fig:3d}
\end{figure}
\begin{table}[h!] 
\centering
\begin{tabular}{cccc}\hline
Star          & Previous  & New  & $\Delta$M \\ 
	      & masses	& masses & 	\\\hline
%	& (Werner \& Herwig 2006) & (this work)	& 	\\ \hline 

H 1504+65            & 0.89  & 0.83  & -0.06 \\
RX J0122.9-7521      & 0.72  & 0.65  & -0.07 \\
RX J2117.1+3412      & 0.70  & 0.72  &  0.02 \\
HE 1429-1209         & 0.67  & 0.66  & -0.01 \\
PG 1520+525          & 0.67  & 0.61  & -0.06 \\
PG 1144+005          & 0.60  & 0.55  & -0.05 \\
Jn 1                 & 0.60  & 0.55  & -0.05 \\
NGC 246              & 0.72  & 0.75  &  0.03 \\
PG 1159-035          & 0.60  & 0.54  & -0.06 \\
NGC 650              & 0.60  & 0.54  & -0.06 \\
Abell 21=Ym29        & 0.58  & 0.53  & -0.05 \\
K 1-16               & 0.58  & 0.54  & -0.04 \\
Longmore 3           & 0.59  & 0.54  & -0.05 \\
PG 1151-029          & 0.60  & 0.58  & -0.02 \\
VV 47                & 0.59  & 0.53  & -0.06 \\
HS 2324+3944         & 0.58  & 0.53  & -0.05 \\
Longmore 4           & 0.65  & 0.63  & -0.02 \\
SDSS J001651.42      & 0.65  & 0.63  & -0.02 \\
SDSS J102327.41      & 0.65  & 0.59  & -0.06 \\
PG 1424+535          & 0.57  & 0.51  & -0.06 \\
HS 1517+7403         & 0.57  & 0.51  & -0.06 \\
Abell 43             & 0.59  & 0.53  & -0.06 \\
NGC 7094             & 0.59  & 0.53  & -0.06 \\
SDSS J075540.94      & 0.62  & 0.58  & -0.04 \\
SDSS J144734.12      & 0.62  & 0.58  & -0.04 \\
SDSS J134341.88      & 0.62  & 0.58  & -0.04 \\
SDSS J093546.53      & 0.62  & 0.58  & -0.04 \\
SDSS J121523.09      & 0.62  & 0.58  & -0.04 \\
HS 0444+0453         & 0.59  & 0.55  & -0.04 \\
IW 1                 & 0.56  & 0.50* & -0.06 \\
Sh 2-68              & 0.55  & 0.50* & -0.06 \\
PG 2131+066          & 0.58  & 0.55  & -0.03 \\
MCT 0130-1937        & 0.60  & 0.54  & -0.06 \\
SDSS J034917.41      & 0.60  & 0.54  & -0.06 \\
PG 1707+427          & 0.59  & 0.53  & -0.06 \\
PG 0122+200          & 0.58  & 0.53  & -0.05 \\
HS 0704+6153         & 0.51  & 0.47* & -0.04 \\\hline

\end{tabular} 
\caption{Masses derived from the new tracks presented in this work by adopting the values of \teff\ and g from Werner \& Herwig (2006). Previous masses, which are based on old post-AGB tracks, have been taken from Werner \& Herwig 2006. Asterisks denote masses that have been derived by linear extrapolation and that are consequently more uncertain than the others. The new mean mass for the PG1159 sample is 0.573\msun, 0.044\msun\ lower than previous mean mass.  All masses are in \msun.}
\label{tab:masitas} 
\end{table}
By adopting the new tracks we have derived the theoretical locus of
the models in the g-\teff-M space. We have done this by adopting a
linear interpolation (i.e. by fitting a plane) between selected points
along the theoretical tracks\footnote{Previous to this we tried a
global least square polinomical fitting but it turned to be
unappropriate.}. Once the theoretical surface on the g-\teff-M space
was derived we have calculated spectroscopical masses by adopting the
\teff-g values presented by Werner \& Herwig (2006). In Fig. \ref{fig:3d} 
we show both the theoretical surface and the derived parameters for
real PG1159 stars. The resulting spectroscopical masses are shown in
Table \ref{tab:masitas}.

\section{Conclusions}

In this paper we have presented full evolutionary calculations
appropriate to post-AGB PG1159 stars. In our calculations, the
complete evolutionary stages of PG1159 progenitors having a wide range
of initial stellar masses have been considered, particularly the
evolution through the born again stage and late thermal pulse (LTP)
that occur following the departure from the AGB.  The location of our
PG1159 tracks in the log g-\teff\ diagram and their comparison with
previous calculations as well as the resulting surface composition
have been discussed at some length. In particular, our models are
generally hotter than previous models of similar masses which do not
display an efficient third dredge up during the TP-AGB, but similar to
recent models that include OV (and in which efficient third dredge up
is found during the TP-AGB). This reinforces the idea that theoretical
log g-\teff\ diagram is sensitive to the previous evolution on the
AGB.

Our results reinforces also the idea that the different abundances of $^{14}$N
observed in the surface of those PG1159 stars with undetected hydrogen 
is an indication that
the progenitors of these stars  would have evolved through  born again 
episodes, where most of the hydrogen content
of the remnant is burnt, or  late thermal pulses where hydrogen
is not burnt but instead diluted to very low surface abundances.

We have also derived isochrones from our PG1159 models aimed at discussing
the presence of planetary nebulae in connection with the location 
of their central stars in the  log g-\teff\ diagram and their corresponding 
evolutionary
ages. In this regard, we infer that the progenitors of some 
PG1159 stars (particularly
NGC 650, VV47, IW,  and PG1520+525) might not have experienced a born again
episode, in agreement with the correlation between the
presence of $^{14}$N and pulsating PG1159s (Dreizler \& Heber 1998).
We also discuss the correlation
between the presence of planetary nebulae and the
$^{14}$N abundance as another indicator that $^{14}$N-rich
objects should come from a born again episode while $^{14}$N-deficient 
should come from a LTP.

 Also, based on the new models presented in this article, we derived
new values for spectroscopical masses. We find that new masses are
systematically lower and consequently at variance with those derived
from asteroseismology.  Although typical uncertainties in
spectroscopical masses, and also uncertainties in asteroseismological
masses (as they are model- and method- dependent, Quirion et al. 2004,
Corsico \& Althaus 2006), prevent us from making concluding remarks,
we think that this difference may be indicating that the microphysics
or the previous evolution (as for example the number of thermal pulses
on the AGB) may be somewhat different than adopted in this work. Both
improved determinations in g and T$_{\rm eff}$ for PG1159 stars and
the use of realistic PG1159 models in asteroseismological analysis are
in order to solve this problem.

Finally, we discuss the possibility that the  He-enriched PG1159 stars
can be represented by very low mass objects, the progenitors of which
have abandoned the AGB shortly before the occurrence of the first
thermal pulse. These progenitors experience their first thermal pulse
as a LTP. We find that moderate mass loss rates operating during the
evolution through red giant region after the LTP could turn these
objects into H-deficient ones, but with high surface helium abundances
and deficiency in nitrogen. In particular, we find that the surface
abundances of HS 0704+6153, HS 1517+7403 and MCT 0130-1937, three
helium enriched objects in the sample of Dreizler \& Heber (1998), can
be reproduced within this scenario.  Also our numerical experiment
shows that this scenario provides a possible explanation for the
existence of O(He) stars of low mass, linking them as the immediate
ancestors of the He-enriched PG1159.

Finally, tabulations of $T_{\rm eff}$, $L_\star$ and $g$ for the tracks
presented in this work are freely available at our URL:{\ttfamily
http://www.fcaglp.unlp.edu.ar/evolgroup }

\begin{acknowledgements}

 We warmly acknowledge our referee (K. Werner) for his
comments. We are very thankful for his suggestions which strongly
improve the original version of the paper. We also want to thank
A. C\'orsico for a careful reading of the manuscript and A. Serenelli
for sending us the post-core-helium-flash 1\msun\ model. Part of this
work has been supported by the Instituto de Astrof\'{\i}sica La
Plata.

\end{acknowledgements}


\begin{thebibliography}{}

\bibitem{A94} Alexander, D. R., \& Ferguson, J. W. 1994, ApJ, 437, 879 
\bibitem{AS05} Althaus, L. G., Serenelli, A. M.,  Panei, J. A., et al. 2005a, 
A\&A, 435, 631
\bibitem{AM05} Althaus, L. G., Miller-Bertolami, M. M.,  C\'orsico, A. H., 
Garc\'{\i}a-Berro, E., \& Gil-Pons, P.  2005b, A\&A, 440, L1
\bibitem{Aetal99} Asplund, M., Lambert, D., Pollaco, D., \& Shetrone, M. 
1999 {A\&A}, 343, {507}
\bibitem{B95} Bl{\" o}cker, T. 1995, {A\&A}, 299,755
\bibitem{B01} Bl{\" o}cker, T. 2001, {ApSS}, 275,  {1}
\bibitem{CA} C\'orsico, A. H. \& Althaus, L. G., 2006, A\&A
to be published
\bibitem{DH98} Dreizler, S. \& Heber, U. 1998, {A\&A}, 334,  {618}
\bibitem{F77} Fujimoto, M. Y.  1977, PASJ, 29, 331
\bibitem{GAS05} Gautschy, A., Althaus, L. G., \& Saio, H. 2005, A\&A, 438, 1013
\bibitem{Hetal05} Hajduk, M., Zijltra, A., Herwig, F., et al. 2005, Science, 308, 231
\bibitem{H00} Herwig, F. 2000, A\&A, 360, 952
\bibitem{H01} Herwig, F. 2001, {ApSS}, 275,  {15}
\bibitem{HBLD99} Herwig, F., Bl{\"o}cker, T., Langer, N., \& Driebe, T. 1999, {A\&A}, 349, {L5}
\bibitem{Hetal06} Herwig, F., Freytag, B., Hueckstaedt, M., Timmes, F. 2006, ApJ preprint doi:10.1086/'501119'
\bibitem{I95} Iben, I. Jr. 1995, Physics Reports, 250, 1
\bibitem{IEA83} Iben, I. Jr., Kaler, J. B., Truran, J. W., \&  Renzini, A.
        1983, ApJ, 264, 605
\bibitem{IM95} Iben, I. Jr.\& MacDonald 1995, Lecture Notes in Physics (Berlin: Springer-Verlag),443, 48
\bibitem{IR96} Iglesias, C. A., \& Rogers, F. J. 1996, ApJ, 464, 943
\bibitem{Ietal95} Iglesias, C. A., Wilson, B. G., Rogers, F. J., Goldstein, W. H., Bar-Shalom, A., \& Oreg, J. 1995,  ApJ, 464, 943
\bibitem{LM03} Lawlor, T. M., \& MacDonald, J. 2003, ApJ, 583, 913 
\bibitem{Metal02} Miksa, S., Deetjen, J. L., Dreizler, S., Kruk, J. W., 
Rauch, T., \& Werner, K. 2002, A\&A, 389, 953 
\bibitem{Metal05} Miller Bertolami, M. M., Althaus, L. G., Serenelli, A. M., 
\& Panei, J. A. 2006, A\&A, in press
\bibitem{PEA86} Quirion P. O., Fontaine G., \& Brassard, P. 2004, ApJ, 610, 436
\bibitem{RDW98}  Rauch, T.,  Dreizler, S., \& Wolff, B. 1998, A\&A, 338, 651
\bibitem{Retal04} Rauch, T., Reiff, E., Werner, K., et al. 2004, astro-ph/0410698
\bibitem{SJ02} Saio, H. \& Jeffery, C. S. 2002, MNRAS, 333, 121
\bibitem{S79} Sch\"onberner, D. 1979, A\&A, 79, 108
\bibitem{SB92} Sch\"onberner, D. \& Bl{\" o}cker, T. 1992, Lecture Notes in Physics (Berlin: Springer-Verlag), 401, 305
\bibitem{W01} Werner, K. 2001, {APSS},  275, {27}
\bibitem{WH06} Werner, K., \& Herwig, F. 2006, PASP, in press
\bibitem{Wetal98} Werner, K., Dreizler, S., Heber, U., Rauch, T. 1998, ASP Conference Series, 135, 130
\bibitem{WF86} Wood, P. R., \& Faulkner, D. J. 1986, {ApJ}, 307, {659}

\end{thebibliography}
\end{document}